\documentclass[superscriptaddress]{revtex4}

\usepackage{graphicx}
\newcommand{\MW}{M_W}
\newcommand{\MZ}{M_Z}
\newcommand{\sweff}{\sin^2\theta_{\mathrm{eff}}}
\newcommand{\gammal}{\Gamma_l}
\newcommand{\mt}{m_t}
\newcommand{\MH}{M_H}

\newcommand{\De}{\Delta}
\newcommand{\de}{\delta}
\newcommand{\msbar}{$\overline{\rm MS}$}
\def\order#1{${\cal O}(#1)$}
\def\ul#1{\underline{#1}}
\newcommand{\al}{\alpha}
\newcommand{\als}{\alpha_s}

\begin{document}

\thispagestyle{empty}
\setcounter{page}{0}
\def\thefootnote{\fnsymbol{footnote}}

\begin{flushright}
BNL--HET--01/34\hfill
ER/40685/969\\
DCPT/01/88\hfill
IPPP/01/44\\
UB-HET--01--07\hfill
UPR--963--T\\
UR--1645\hfill
hep-ph/0111314\\
\end{flushright}

\vspace{1cm}

\begin{center}

{\Large\sc {\bf Theoretical and Experimental Status of the indirect}}

\vspace*{0.4cm} 

{\Large\sc {\bf Higgs Boson Mass Determination in the 
Standard Model$^{\ddag\ddag}$}}

\vspace{1cm}

{\sc 
U.~Baur$^{\,1,^*}$%
, R.~Clare$^{\,2,\dag}$%
, J.~Erler$^{\,3,\ddag}$%
, S.~Heinemeyer$^{\,4,\S}$%
,\\[.5em] D.~Wackeroth$^{\,5,\P}$%
, G.~Weiglein$^{\,6,^{**}}$%
\vspace{0.5cm}
 and D.R.~Wood$^{\,7,\dag\dag}$%
}

\vspace*{0.5cm}
 
$^1$ State University of New York at Buffalo, Buffalo, NY 14260, USA

\vspace*{0.1cm}

$^2$ University of California, Riverside, CA 92521, USA

\vspace*{0.1cm}

$^3$ Dept.\ of Physics and Astronomy, University of Pennsylvania, 
     PA 19146, USA

\vspace*{0.1cm}

$^4$ HET Physics Dept., Brookhaven Natl.\ Lab., NY 11973, USA

\vspace*{0.1cm}

$^5$ University of Rochester, Rochester, NY 14627, USA

\vspace*{0.1cm}

$^6$ Institute for Particle Physics Phenomenology, Durham, UK

\vspace*{0.1cm}

$^7$ Northeastern University, Boston, MA 02115, USA

\end{center}

\vspace*{0.4cm}

\begin{center}
{\large\bf Abstract}
\end{center}

The impact of theoretical and experimental uncertainties on the
indirect determination of the Higgs boson mass, $\MH$, in the Standard
Model (SM) is discussed. Special emphasis is put on the electroweak
precision observables $\MW$ (the $W$~boson mass) and $\sweff$ (the
effective leptonic mixing angle). The current uncertainties of the
theoretical predictions for $\MW$ and $\sweff$ due to
missing higher order corrections are conservatively estimated to 
$\de\MW \approx 7$~MeV and $\de\sweff \approx 7 \times 10^{-5}$.
Expectations and necessary theoretical improvements
for future colliders are explored. Results for the indirect $\MH$
determination are presented based
on the present experimental and theoretical precisions as well as on
improvements corresponding to the prospective situation at future
colliders. The treatment of the different future colliders is done in a
uniform way in order to allow for a direct comparison of the
accuracies that can be reached. Taking all experimental, theoretical, and
parametrical uncertainties into account, 
a current upper bound on $\MH$ of $\sim 200$~GeV is obtained. Furthermore
we find in a conservative approach that a Linear Collider
with GigaZ capabilities can achieve a relative precision of about~8\%
(or better) in the indirect determination of $\MH$.
 

\vfill

\begin{picture}(200,0.2)
\put(0,0){\line(1,0){200}}
\end{picture}

$^{\ddag\ddag}$ Contribution to the P1-WG1 report, 
``Workshop on the Future of Particle Physics'', Snowmass, Colorado, 
\mbox{}\hspace{6mm} USA, July 2001.

\vspace{0.5em}
$^*$ email: baur@ubhex.physics.buffalo.edu

$^\dag$ email: robert.clare@ucr.edu

$^\ddag$ email: erler@langacker.hep.upenn.edu

$^\S$ email: Sven.Heinemeyer@bnl.gov

$^\P$ email: dow@pas.rochester.edu

$^{**}$ email: Georg.Weiglein@durham.ac.uk

$^{\dag\dag}$ email: darien@neu.edu

\newpage

\bibliographystyle{revtex}

\title{Theoretical and Experimental Status of the Indirect Higgs Boson
Mass Determination in the Standard Model}

\author{U. Baur}
\email[]{baur@ubhex.physics.buffalo.edu}
\affiliation{State University of New York at Buffalo, Buffalo, NY 14260,
USA}

\author{R. Clare}
\email[]{robert.clare@ucr.edu}
\affiliation{University of California, Riverside, CA 92521}

\author{J. Erler}
\email[]{erler@ginger.hep.upenn.edu}
\affiliation{Dept.\ of Physics and Astronomy, University of Pennsylvania,
PA 19146, USA}

\author{S. Heinemeyer}
\email[]{Sven.Heinemeyer@bnl.gov}
\affiliation{HET Physics Dept., Brookhaven Natl.\ Lab., NY 11973, USA}

\author{D.~Wackeroth}
\email[]{dow@pas.rochester.edu}
\affiliation{University of Rochester, Rochester, NY 14627, USA}

\author{G. Weiglein}
\email[]{Georg.Weiglein@durham.ac.uk}
\affiliation{Institute for Particle Physics Phenomenology, Durham, UK}

\author{D.R.~Wood}
\email[]{darien@neu.edu}
\affiliation{Northeastern University, Boston, MA 02115, USA}

\begin{abstract}
The impact of theoretical and experimental uncertainties on the
indirect determination of the Higgs boson mass, $\MH$, in the Standard
Model (SM) is discussed. Special emphasis is put on the electroweak
precision observables $\MW$ (the $W$~boson mass) and $\sweff$ (the
effective leptonic mixing angle). The current uncertainties of the
theoretical predictions for $\MW$ and $\sweff$ due to
missing higher order corrections are conservatively estimated to 
$\de\MW \approx 7$~MeV and $\de\sweff \approx 7 \times 10^{-5}$.
Expectations and necessary theoretical improvements
for future colliders are explored. Results for the indirect $\MH$
determination are presented based
on the present experimental and theoretical precisions as well as on
improvements corresponding to the prospective situation at future
colliders. The treatment of the different future colliders is done in a
uniform way in order to allow for a direct comparison of the
accuracies that can be reached. Taking all experimental, theoretical, and
parametric uncertainties into account, 
a current upper bound on $\MH$ of $\sim 200$~GeV is obtained. Furthermore
we find in a conservative approach that a Linear Collider
with GigaZ capabilities can achieve a relative precision of about 8\%
(or better) in the indirect determination of $\MH$.
\end{abstract}

\maketitle


\section{Introduction}
\label{sec:mhindirect}

In this contribution we address the status and possible future
developments in the measurements of and the theoretical predictions for
the most important electroweak precision observables. We estimate
their precision from upcoming and proposed accelerator experiments.
In all cases we quote uncertainties which we believe to be realistically
achievable, not excluding even greater precisions.  As a result of
imposing similar standards in all cases, our quoted uncertainties 
should be directly comparable.  Similarly, we attempt to anticipate which 
improvements can be expected in the theoretical predictions for 
the observables. Again, we believe that our estimates can be 
realistically achieved with a dedicated effort and allow some leeway
for even more precise calculations.

\begin{figure}[t]
\includegraphics[height=10cm]{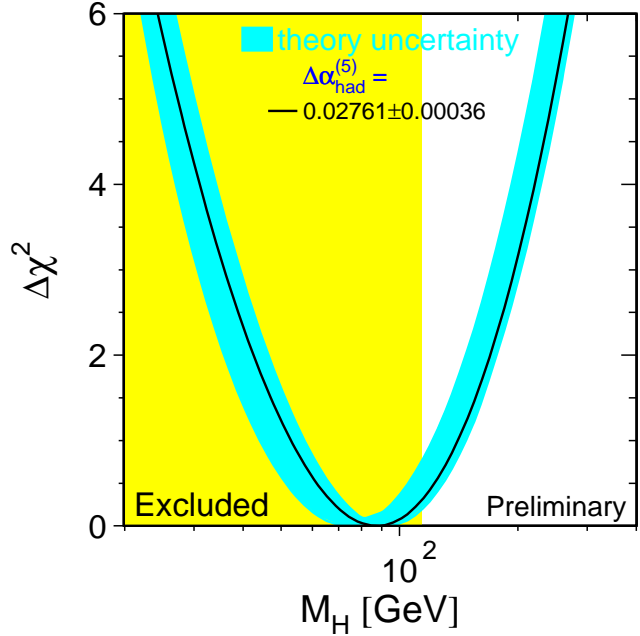}
\caption{$\De\chi^2=\chi^2 - \chi^2_{\rm min}$ from a global fit to all
available data~\cite{ewwg} as a function of the SM Higgs boson mass,
$M_H$. The width of
the ``Blue Band'' indicates the effect of ``intrinsic'' uncertainties
from unknown higher order corrections (see text). The yellow region is
excluded by direct Higgs searches at LEP2~\cite{lephiggs}.}
\label{blueband}
\end{figure}

Within the SM, the mass of the Higgs boson, $M_H$, can be constrained 
indirectly
with the help of electroweak precision observables (EWPO).
As a result of a global analysis, Fig.~\ref{blueband}~\cite{ewwg}
shows $\De\chi^2 \equiv \chi^2 - \chi^2_{\rm min}$ as
an approximately quadratic function of $\log M_H$. Therefore, the 95\%
CL upper
limit can be approximated by $\De\chi^2 = 2.71$, corresponding to a
95\% CL upper bound of $M_H < 196$~GeV at present.

\smallskip

Among the experimental measurements of EWPO which are used in
global fits, the $W$ boson mass, $\MW$, and the effective leptonic
weak mixing angle, $\sweff$, have the largest impact on the
extracted value of $\MH$. Although the current relative precision
of $\MW$ is better by a factor of 1.8~compared to $\sweff$, the
latter is the most relevant parameter for the indirect $\MH$
determination due to its more pronounced dependence on the Higgs
mass. For equal relative experimental precisions, it yields a
3.1~times higher sensitivity (for $\MH$ around 115~GeV). Other
observables include the leptonic $Z$~boson width, $\gammal$; the
mass and width of the $Z$~boson, $\MZ$ and $\Gamma_Z$; the peak
hadronic cross section of the $Z$~boson, $\sigma^0_{\rm had}$;
EWPO from deep inelastic neutrino scattering; and others.
Furthermore the top quark mass, $\mt$, enters in the global fit;
its value and its error have a strong impact on the extracted
$\MH$ value.

The precision of the fit results depends on the experimental uncertainties of
the measured values of the EWPO and the theoretical uncertainties of
their predictions.
When discussing these uncertainties, one has to take into account that
most of the EWPO, for example $M_Z$, $\MW$ and $\sweff$, are not
directly measurable quantities, but are related to measured
cross-sections and asymmetries by a deconvolution or unfolding
procedure. They are therefore often called ``pseudo-observables'', in
order to distinguish them from directly measured ``primordial''
observables. The unfolding procedure is in general affected by
theoretical uncertainties (and a certain degree of model dependence),
which enter the systematic experimental error of the pseudo-observables.
We will refer to this kind of theoretical uncertainties as
{\em primordial\/} theoretical uncertainties in the following. A second
kind of theoretical uncertainty arises in the prediction for
pseudo-observables, e.g.\ $\MW$ and $\sweff$, in terms of the 
chosen input parameters within a certain model,
e.g.\ the Standard Model or the Minimal Supersymmetric Standard Model. 
We use the phrase {\em intrinsic\/} uncertainties for the ones arising
from unknown higher-order corrections in the perturbative expansion,
as well as for other uncertainties arising from computational limitations.
Finally, {\em parametric\/} errors originate from the limited experimental 
precision on the input parameters. The effect of the intrinsic uncertainties 
is indicated by the width of the ``Blue Band'' in Fig.~\ref{blueband}.

The SM predictions for the EWPO are calculated in terms of a small set
of input parameters:
$M_Z, G_{\mu}, \alpha (M_Z), m_\ell, m_q,
\mt, \MH$, and $\alpha_s(M_Z)$.  The fine structure constant, $\alpha(0)$, 
the $Z$ boson mass, $\MZ$, the lepton masses,
$m_\ell$, and the Fermi constant, $G_{\mu}$, are currently the most
precisely measured input parameters~\cite{pdg}, and their errors have
negligible effects on the fit results~\cite{Erler:1995fz,lepewwg,pgl}.
The dominant uncertainties presently arise from the experimental error on
the top quark mass, $\mt = 174.3 \pm 5.1$~GeV~\cite{pdg}, the hadronic
contribution to the fine structure constant,
$\De\alpha_{\rm had}$~\cite{jegerlehner,deltaalpha}
(the value used in Fig.~\ref{blueband} is from Ref.~\cite{deltaalpha1}),
as well as $\MH$. $\alpha_s(M_Z)$ is constrained mainly by
$\Gamma_Z$, $R_l$, and $\sigma^0_{\rm had}$, with little theoretical
uncertainty as long as one ignores the possibility of large new physics
effects.

In practice, both EWPO and input parameters are used as constraints
in the fits subject to their experimental uncertainties (which, as
explained above, contain the primordial theoretical uncertainties
related to extraction of the EWPO). The only distinction
is that the input parameters are treated as fit parameters, and the EWPO
are computed in terms of these.  For example, $m_t$ which appears
only in loops is chosen as input. Moreover, one usually prefers to compute
less precise quantities in terms of more precise ones.  The fit results
are insensitive to these choices.

Table~\ref{expfuture} summarizes the current status of the
experimental uncertainties and the precision one expects to achieve at
future colliders for the most relevant EWPO, $\MW$ and $\sweff$, and the top
quark mass, together with the expected experimental error on $\MH$,
assuming the SM Higgs
boson has been discovered with $\MH \approx 115$~GeV. The entries in the table
attempt to represent the combined results of all detectors and channels at
a given collider, taking into account correlated systematic uncertainties.

\begin{table}[bht]
\caption{
The expected experimental uncertainties (including theory errors for the
experimental extraction, i.e.\ the primordial uncertainties, see text)
at various colliders are summarized
for $\sweff$, $\MW$, $\mt$, and $\MH$ (the latter assuming $\MH = 115$~GeV).
Each column represents the combined results of all detectors and
channels at a given collider, taking into account correlated
systematic uncertainties.
\newline
Run~IIA refers to an integrated luminosity of 2~fb$^{-1}$ (per detector)
collected
at the Tevatron with the Main Injector, while Run~IIB (IIB$^*$) assumes
the accumulation of 15 (30)~fb$^{-1}$. The numbers for $\sweff$ are
obtained by scaling  (see Ref.~\cite{talkmschmitt}) the uncertainties of 
Run~I~\cite{sw2effcdf} to the quoted integrated luminosities. 
A detailed analysis~\cite{run2prec} has shown that the uncertainties
for $\sweff$ approximately scale with $1/\sqrt{\cal L}$. 
Earlier estimates~\cite{run2prec} were based on the approximation of a linear
relationship between the forward-backward
asymmetry, $A_{\rm FB}$, and $\sweff$. The numbers given here have
additionally been
corrected to reflect the full tree level relation between $A_{\rm FB}$ and
$\sweff$. The values for $\MW$ are
taken from Ref.~\cite{talkmschmitt}, while the $\MH$ uncertainty is
from Ref.~\cite{gunion}.
\newline
The upper end of the $\de\sweff$ range (used for the fits in
Table~\ref{indirectmh}) at the LHC corresponds to the statistical
uncertainty which can be obtained 
in one year of running at high luminosity (100~fb$^{-1}$) after
combining the $e$ and $\mu$ channels and the two experiments~\cite{haywood}.
Systematic uncertainties and cross correlations have been ignored in this
estimate.  However, one can gain considerable leeway by accumulating data over
several years. Moreover, one may be able to increase the pseudorapidity range
(see text) potentially allowing even greater precision.  This is indicated in
the range. The uncertainty of 
15~MeV quoted for $\MW$ at the LHC is challenging but should be feasible
due to the enormous statistics~\cite{haywood}. For the Higgs boson mass
uncertainties at the LHC, see Ref.~\cite{lhchiggs}.
\newline
LC denotes a linear collider operating at $\sqrt{s}=500$~GeV. The uncertainty
quoted for $\MW$ is based on an integrated luminosity of
500~fb$^{-1}$~\cite{klausyboy}. 
(The entry in parentheses assumes a fixed target polarized M\o ller scattering
experiment using the $e^-$ beam~\cite{kumar,baurdem}, thus corresponding
to an effective mixing angle at a scale of 
${\cal O}(0.5$~GeV). It is not used in the fits.)
\newline
GigaZ collectively denotes an LC operating at $\sqrt{s}=M_Z$ or
$\sqrt{s}\approx 2 \MW$ with a luminosity of 
${\cal L} \approx 5 \times 10^{33}\,{\rm cm}^{-2}\,{\rm s}^{-1}$.
The GigaZ error for $\MW$ combines the 5.2~MeV
experimental error~\cite{wilson} (requiring about one year of
running) with beam energy and theory uncertainties
(see text) which for definiteness we assume close to 3 MeV each
(which is challenging). The determination of $\sweff$ with the quoted
precision at GigaZ can be
performed in 50-100 days of running, see Ref.~\cite{moenig} for
details. $\de\MH$ at the LC/GigaZ is discussed in detail in 
Refs.~\cite{teslatdr,orangereport,acfa}. 
\newline
$\de\mt$ from the Tevatron~\cite{run2prec,talkmschmitt}
and the LHC~\cite{beneke} is the uncertainty in
the top pole mass. We included an irreducible uncertainty of order
$\Lambda_{\rm QCD} \sim 0.5$~GeV from non-perturbative and renormalon
ambiguities. The precision listed for GigaZ and the LC is for the \msbar\ top
mass, see Refs.~\cite{tttheo,ttexp}.
The relatively smaller uncertainty at GigaZ compared to the LC is due to
the higher precision in $\alpha_s$ (from other GigaZ observables)
which affects the extraction of $m_t$.}
\vspace{10pt}
\label{expfuture}
\renewcommand{\arraystretch}{1.5}
\begin{tabular}{|c||c||c|c|c|c||c|c|}
\cline{2-8} \multicolumn{1}{c||}{}
& now & Tev.\ Run~IIA & Run~IIB & Run~IIB$^*$ & LHC & ~LC~  & GigaZ \\
\hline\hline
$\de\sweff(\times 10^5)$ & 17   & 78   & 29   & 20   & 14--20 & (6)  & 1.3  \\
\hline
$\de\MW$ [MeV]           & 33   & 27   & 16   & 12   & 15   & 10   & 7      \\
\hline
$\de\mt$ [GeV]           &  5.1 &  2.7 &  1.4 &  1.3 &  1.0 &  0.2 & 0.13   \\
\hline
$\de\MH$ [MeV]            &  --- &  --- &
                 \multicolumn{2}{c|}{${\cal O}(2000)$} &  100 &   50 &   50 \\
\hline
\end{tabular}
\renewcommand{\arraystretch}{1}
\end{table}

The current measurement of $\sweff$ is dominated by the left-right
asymmetry from SLD and the $b$-quark forward-backward asymmetry from
LEP1.  In the future, constraints are expected to come from precise
measurements of the forward-backward asymmetry in
$p\overline{p}(pp)\to Z+X\to \ell^+\ell^- +X$ at the Tevatron and the LHC,
and from a possible linear collider (LC).  At the LC, $\sweff$ can be
determined from left-right asymmetries when operating at the $Z$~peak
(GigaZ). Another effective mixing angle
(at a much lower energy scale) can also be measured in fixed-target 
M\o ller scattering. We do not consider this measurement, which would
provide additional input for precision tests of the SM, in more detail here.
In both cases, polarized beams are needed.

The current precision of $\MW$ is dominated by the direct mass
reconstruction of $W$-pair events at
LEP2. Transverse-mass fits from
Run~I at the Tevatron and data from threshold scans at LEP2 also
contribute significantly but carry less statistical weight. Future LHC
and Tevatron estimates are based on fits to the transverse mass distribution,
the lepton transverse momentum distribution, and the $W/Z$ transverse mass
ratio~\cite{run2prec}. The LC estimate in the continuum is based on
the direct mass reconstruction of $W$-pair events, similar to the LEP2
analysis~\cite{klausyboy}.  The GigaZ projection assumes a dedicated
threshold scan,
which requires that the knowledge of the absolute beam energy can be
controlled better than 2.5~MeV~\cite{wilson}.

The determination of the top quark mass at the Tevatron (present and
future) and at the LHC is based on kinematic reconstruction, and thus
represents a measurement of the pole
mass~\cite{run2prec,talkmschmitt,beneke}.  At the LC, $\mt$ can be
determined from an energy scan near the $t\overline{t}$ production
threshold. The extracted value can be chosen to correspond to a suitably
defined threshold mass or another short distance mass such as the \msbar\
mass~\cite{tttheo,ttexp}.  The threshold analysis gives correlated
measurements of $\alpha_s$
and $\mt$, and the last entry in Table~\ref{expfuture} represents
the combination of the threshold scan with the precise $\alpha_s$
determination from GigaZ~\cite{alphas,jenssvengigaz}.  A recent
calculation~\cite{mtnew} may allow for an even better determination
of $\mt$ up to $\sim 50$~MeV.

In Section \ref{sec:currentmh} we summarize the current status of the
intrinsic and parametric uncertainties in the predictions for the most
relevant EWPO and  
analyze their impact on the current prediction for $M_H$ within the SM. 
In Section \ref{sec:futuremh} we
discuss necessary improvements of primordial theoretical uncertainties
which are required for the extraction of $\MW$, $\sweff$, and $m_t$ at
future colliders with a precision as envisaged in Table~\ref{expfuture}.
We also analyze the necessary improvements in the predictions of the
EWPO in order to match the experimental precisions. Based on
estimates of prospective improvements of the experimental and
theoretical uncertainties, we study the accuracy which can be achieved in the
indirect determination of $\MH$ at future colliders.


\section{Current theoretical uncertainties of EWPO and the
prediction of $\MH$}
\label{sec:currentmh}

The dominant parametric uncertainty of the EWPO presently arises from
the experimental error of the top quark mass, $\de\mt=5.1$~GeV.
This error induces a parametric uncertainty of 32~MeV and
$16 \times 10^{-5}$ in
the $W$~mass and the leptonic effective mixing angle, respectively.
The corresponding errors from the uncertainty in
$\De\alpha_{\rm had}$, $\de\De\alpha_{\rm had} = 0.0002$, are 3.7~MeV and
$7 \times 10^{-5}$.  Furthermore, the imperfect knowledge of the
strong coupling
constant, $\de\alpha_s(M_Z) = 0.0028$, introduces uncertainties of 2~MeV
and $3.5 \times 10^{-5}$ and also an uncertainty in $\De\al_{\rm had}$
of about $\de\De\al_{\rm had} = 0.0001$.
While the uncertainty induced by the top quark
mass is about as large as the present experimental error of $\MW$
and $\sweff$, the parametric uncertainties induced by the errors of
$\De\alpha_{\rm had}$ and $\alpha_s(M_Z)$ are already smaller than the
prospective experimental errors on $\MW$ and $\sweff$ at the Tevatron
and the LHC (see Table~\ref{expfuture}). On the other hand, the
accuracies reachable at GigaZ will clearly require a significantly
improved experimental precision not only of $\mt$ (see
Table~\ref{expfuture}), but also of $\De\alpha_{\rm had}$ and
$\alpha_s(M_Z)$. An improved determination of $\alpha_s(M_Z)$ with
little theoretical uncertainty is, in fact, expected from GigaZ
itself~\cite{jenssvengigaz,alphas}.

Concerning the intrinsic uncertainties of the EWPO from unknown higher
orders, recent progress has been made for the prediction of $\MW$ by
the inclusion of the full fermionic two-loop
corrections~\cite{weiglein}, superseding the previous expansions in
$\mt^2/\MW^2$. Since this expansion yielded similar values (with the
same sign) for the $\mt^4/\MW^4$ and the $\mt^2/\MW^2$ terms (casting
some doubt on the convergence), the full fermionic two-loop corrections
constitute an important step towards a very precise $\MW$ prediction.
The difference between the expansion calculation and the full result can
reach up to about 4~MeV, depending on $\MH$. The only missing two-loop
corrections to $\MW$ are the pure
bosonic contributions. The $\MH$ dependence of the bosonic two-loop
contributions to $\MW$ has recently been evaluated~\cite{MWMHdep2},
indicating corrections of \order{1~{\rm MeV}}. For $\sweff$ the
situation is slightly worse, since a result for the full fermionic
two-loop corrections is not yet available, and one has to rely on the
expansion in powers of $m^2_t/\MW^2$~\cite{sweffexpansion}.
Beyond two-loop order, the results for the pure fermion-loop
contributions (incorporating in particular the leading terms in
$\Delta\alpha$ and $\Delta\rho$) are known up to the four-loop
order~\cite{fermloop}. Furthermore, the QCD corrections of
${\cal O}(\al\als^2)$ are known~\cite{chetyrkin,chetstein}. More
recently, also the leading three-loop terms
of ${\cal O}(G_F^3 m_t^6)$ and ${\cal O}(G_F^2 \als m_t^4)$, which enter
via the quantity $\Delta\rho$, have been calculated in the limit of
vanishing Higgs boson mass. The results have been found to be quite
small, which is familiar from the $\MH = 0$ limit of the
${\cal O}(G_F^2 m_t^4)$ result~\cite{vanderbij}. In the latter case,
the extension to finite values of $\MH$ and the inclusion of subleading
terms led to an increase in the numerical result by a factor of up to 20.

In order to quantify the remaining intrinsic uncertainties of the EWPO,
one has to perform estimates of the possible size of uncalculated
higher-order corrections.
The results of calculations based on different renormalization
schemes or on different prescriptions for incorporating non-leading
contributions in resummed or expanded form differ from each other
by higher-order corrections. One way of estimating the size of unknown
higher-order corrections is thus to compare the results for the
prediction of the EWPO from different codes in which the same
corrections are organized in a somewhat different way. A detailed
description of different ``options'' used in this comparison can be
found in Ref.~\cite{yellowbook1995} and an update in Ref.~\cite{bardin1999}.
This prescription may lead to an underestimate of the theoretical
uncertainty if at an uncalculated order a new source of potentially
large corrections (e.g.\ a certain enhancement factor) sets in. In
general, it is not easy to quantify how large the variety of different
codes and different ``options'' should be in order to obtain a
reasonable estimate of the theoretical uncertainty.

The ``Blue Band'' in Fig.~\ref{blueband} is obtained according to the
prescription described above, using the codes {\tt ZFITTER}~\cite{ZFITTER} and
{\tt TOPAZ0}~\cite{TOPAZ0}. At present, the theoretical uncertainty
represented by
the width of the ``Blue Band'' mainly arises from the
intrinsic uncertainties in the prediction for $\sweff$, while the
intrinsic uncertainty in the prediction for $\MW$, being significantly
smaller than the experimental error, is less important. A shift in the
prediction for $\MW$, on the other hand, induces a shift in $\sweff$
according to
\begin{equation}
\sweff = \left(1 - \frac{\MW^2}{\MZ^2} \right) \kappa(\MW^2) ,
\label{eq:sweffshift}
\end{equation}
where $\kappa$ is a calculable function in the SM.
While the shift in $\MW$ induced by going from the result of the
expansion in powers of $m^2_t/\MW^2$ to the result of the full fermionic
two-loop corrections is known, the corresponding result for
$\kappa(\MW^2)$ is still missing. 
The effect of inserting the new result for $\MW$ in
Eq.~(\ref{eq:sweffshift}), which amounts to an upward shift of about
$8 \times 10^{-5}$ in $\sweff$ (for $\MH \approx 115$~GeV), has been
(conservatively) treated as  
a theoretical uncertainty in the ``Blue Band'' of Fig.~\ref{blueband}.

Other (related) methods to estimate the size of missing higher order
corrections are to vary the renormalization scales and schemes. While
these methods usually give an order of magnitude estimate and a lower
bound on the uncertainty, they can lead to underestimates whenever there
are sizeable but scheme- and scale-invariant contributions.  For example,
the lowest order flavor singlet contribution to $Z$ decay, a separately
gauge invariant and finite set of corrections, cannot be estimated by
scale variations of the non-singlet contribution or by using different
``options'' for resumming non-leading contributions
in computer codes. 

In the following we use a simple minded, but rather robust and, in the
past, quite successful method for estimating the uncertainties from
unknown higher orders~\cite{GAPP}.
The idea is to collect all relevant enhancement and suppression
factors and setting the remaining coefficient (from the actual loop
integrals) to unity. If, in a given order, terms with different group
theory factors contribute, one can often choose the largest one as
an estimate for the uncertainty.
Our results are summarized in Table~\ref{higherorderunc}.
They are in good agreement with the estimates of the current
uncertainties of $\MW$ and $\sweff$ performed in
Refs.~\cite{gambunc,sirlunc,MWMHdep2,freitunc}.

\begin{table}[thb]
\caption{
Theoretical uncertainties from unknown higher-order corrections to
$\sweff$ and  $\MW$. $\hat{s}$ denotes the \msbar\ mixing angle,
$N=12$ is the number of  fermion doublets in the SM, $C_F = 4/3$ and
$C_A = 3$ are QCD factors, and  $N_C = 3$ is the number of colors.
The corrections in the upper part of the table are assumed to enter the
predictions in the same way as $\Delta\alpha$ (only the leading top quark
correction of \order{\al\als^2} enters via $\De\rho$), while the ones in the
lower part are assumed to enter via $\Delta\rho$.
The fermionic contributions of ${\cal O}(\alpha^2)$ refer to the
non-leading terms beyond the next-to-leading term of the expansion in
powers of $m_t^2/\MW^2$. The uncertainty in $\sweff$ has been
estimated from the known correction to $\MW$ using
Eq.~(\ref{eq:sweffshift}) (see text).
The ${\cal O}(\al\als^2)$ corrections,
which are completely known both for $\MW$ and
$\sweff$, are included in the table for completeness.
However, the light fermion corrections are not yet included in all codes
currently used for performing electroweak fits (and have not been
published yet as an independent explicit formula);
our error estimate for $\MW$ and $\sweff$ corresponds to 
$\pm 1.7$~MeV and $\pm 3.3\times 10^{-5}$, respectively.
In order to estimate effects of finite $\MH$ and subleading terms
in the lower part of the table, we have taken the average of the
individual coefficients of the result in the limit
$\MH = 0$~\cite{Chetyrkin3} (which in this limit conspire to yield a
small answer), resulting in the numerical prefactors there.
}
\vspace{10pt}
\label{higherorderunc}
\renewcommand{\arraystretch}{1.5}
\begin{tabular}{|c|c|c|c|c|c|}
\hline
 order & sector & estimate & size ($\times 10^{5}$) & $\MW$[MeV] & $\sweff$ ($\times 10^{5}$) \\
\hline \hline
$\al^2$ & fermionic & $N(\al/4\pi\hat{s}^2)^2$ & 8.7 & complete~\cite{weiglein} & 4.1 \\
$\al^2$ & bosonic & $(\al/\pi\hat{s}^2)^2$ & 11.6 & 2.1 & 4.1 \\
$\al\als^2$ & top-bottom doublet & $N_C C_F C_A \al\als^2/4\pi^3\hat{s}^2$ & 4.7 & complete~\cite{chetyrkin} & complete~\cite{chetyrkin} \\
$\al\als^2$ & light doublets & $2\; N_C C_F C_A
\al\als^2/4\pi^3\hat{s}^2$ & 9.4 & complete~\cite{chetstein} &
complete~\cite{chetstein} \\
\hline
$\al^3\mt^6$ & heavy top & $5.3\; N_C^2(\al\mt^2/4\pi\hat{s}^2\MW^2)^3$ & 7.0 & 4.1 & 2.3 \\
$\al^3\mt^6$ & heavy top & $3.3\; N_C(\al\mt^2/4\pi\hat{s}^2\MW^2)^3$ & 1.5 & 0.9 & 0.5 \\
$\al^2\als\mt^4$ & heavy top & $3.9\; N_C C_F\al^2\als\mt^4/16\pi^3\hat{s}^4\MW^4$ & 7.8 & 4.5 & 2.5 \\
$\al\als^3\mt^2$ & heavy top &
 $N_C C_F C_A^2 \al\als^3\mt^2/4\pi^4\hat{s}^2\MW^2$ & 2.3 & 1.3 & 0.8
 \\
\hline\hline
& total & & & 7 & 7 \\
\hline
\end{tabular}
\renewcommand{\arraystretch}{1}
\end{table}

\bigskip
We have performed a global fit to all data in the Standard Model based
on the present experimental and parametric uncertainties and using the
estimates of Table~\ref{higherorderunc} for the intrinsic theoretical
uncertainties from unknown higher-order corrections. For the
theoretical predictions the program {\tt GAPP}~\cite{GAPP} has been used.
In contrast to the fit in Fig.~\ref{blueband}, where the theory
uncertainties are represented by the width of the blue band,
we have added theoretical and experimental errors in quadrature. As a
result we find
\begin{equation}
\MH = 97^{+ 53}_{- 36} {\rm ~GeV},
\label{eq:presentfit}
\end{equation}
and a 95\% CL upper bound of $\MH < 194$~GeV.
These numbers are very close to the result of the fit in
Fig.~\ref{blueband}~\cite{ewwg}.

Concerning the interpretation of the fit result, it should be kept in
mind that it is based on the assumption that the Standard Model
provides the correct description of the experimental measurements.
This means, in particular, that the resulting bound on $\MH$ does not
reflect the quality of the fit, i.e.\ it does not
contain information about how well the SM actually describes
the data.


\section{Future indirect determinations of $\MH$}
\label{sec:futuremh}

For the analysis in this section, we
anticipate that in the future the currently missing corrections
indicated in the upper part of Table~\ref{higherorderunc} will
become available,
and that the uncertainties listed in the lower part will be reduced
by a factor of two.

In the following we will discuss the anticipated future
experimental precisions of the EWPO reachable at the next
generation of colliders as given in Table~\ref{expfuture} in
view of necessary improvements of the primordial theoretical
uncertainties. In each case we also
investigate whether the prospective parametric and
intrinsic theoretical uncertainties of the EWPO will be sufficiently
under control in order to match the projected experimental precision.

\begin{itemize}
\item \ul{Tevatron Run~IIA (2~${\rm fb}^{-1}$/experiment):}\\
In order
to measure the $W$ mass with the precision anticipated for Run~IIA, it
is necessary to take into account QCD and electroweak radiative
corrections to $W$ production. In particular, the understanding of QED
radiative corrections which shift the $W$ mass extracted from data by
50~--~150~MeV~\cite{cdfwmass,cdfwmass2,d0wmass,d0wmass2} is crucial
for a precision $W$ mass measurement. The determination of the $W$
mass in a hadron collider environment requires a simultaneous
precision measurement of the $Z$ boson mass, $M_Z$, and width,
$\Gamma_Z$. When compared to the value measured at LEP1, the two
quantities help to accurately calibrate detector components. It is
therefore also necessary to understand the EW corrections to $Z$ boson
production in hadronic collisions. In order to properly calibrate the
$Z$ boson mass and width using the available LEP1 data, it is
desirable to obtain the predictions for the $Z$ observables in
hadronic collisions with an accuracy which is comparable with that of
the theoretical
input which has been used to extract $M_Z$ and $\Gamma_Z$ at
LEP1. During the last three years, results for the full ${\cal O}(\alpha)$
corrections to $W$~\cite{eww,eww2} and $Z$ boson
production~\cite{ewz,ewz2} became available. The remaining
uncertainties from unknown higher order corrections have been
estimated to be of ${\cal O}(5$~MeV)~\cite{run2prec}.

QCD corrections only indirectly influence the $W$ mass determination
via the angular distribution of the decay
lepton~\cite{run2prec}. However, in order to correctly reconstruct the
transverse momentum of the neutrino, it is crucial to accurately
predict the $W$ transverse momentum distribution. This is achieved
using the observed $Z$ boson $p_T$ distribution together with
calculations~\cite{resbos,resbos2,resbos3}
which resum the QCD corrections to the $W$
and $Z$ $p_T$ distributions to all orders, and a parameterization of
non-perturbative effects at small $p_T$~\cite{landry}. The systematic
uncertainties due to the knowledge of the $p_T^W$ distribution are
estimated to be $\de\MW\approx 5$~MeV in Run~IIA. Incomplete knowledge
of the parton distribution functions (PDFs)
will contribute an uncertainty of similar size~\cite{run2prec}.

The effective leptonic mixing angle is expected to be measured from
the precise
determination of the forward-backward asymmetry in $p\bar p\ \to Z + X
\to l^+l^- + X$ at the $Z$ peak.  The main theoretical uncertainty
originates from the incomplete knowledge of the
PDFs~\cite{run2prec}.

In summary, for the precision $\MW$ and $\sweff$ measurements
envisioned at Run~IIA, no further improvements in the
primordial uncertainties of the theoretical predictions
for $W$ and $Z$ observables are needed.

As in Run I, the top quark mass measurement is mainly based
on the direct kinematic reconstruction of the $t \bar t$ events
in the lepton$+$jets channel,
$p\bar p\to t\bar t \to W^+W^-b\bar b\to l^+\nu q \bar q' b\bar b$.
This channel provides a large and clean sample for mass reconstruction,
resulting in a measurement of the top pole mass, and thus
does not require additional theoretical input. In Run~II the
uncertainty in the lepton$+$jets channel
will be dominated by systematic effects, which are
largely dominated by the uncertainty on the jet energy scale and the
modeling of QCD radiation in top events.

The improvement in the experimental determination of $m_t$ at
Run~IIA will reduce the parametric theoretical uncertainties of
$\MW$ and $\sweff$ to values below the experimental errors of
these observables. Similarly, the present values of the
intrinsic theoretical uncertainties of $\MW$ and $\sweff$ (see
Table~\ref{higherorderunc}) are smaller than the envisaged
experimental errors (see Table~\ref{expfuture}). On the other
hand, an improvement of the theoretical prediction of $\sweff$,
in particular the inclusion of the missing corrections of
${\cal O}(\alpha^2)$, would lead to a significant reduction of
the width of the ``Blue Band'' shown in Fig.~\ref{blueband}. It is not
obvious, however, that full two-loop results for $\MW$ and
$\sweff$ will become available already at the time scale of
Run~IIA, which is expected to be completed within the next two
to three years.

\item \ul{Tevatron Run IIB (15 ${\rm fb}^{-1}$/experiment):}\\
The main contribution to the shift in $\MW$ induced by the QED
corrections originates from final state photon radiation. An explicit
calculation of real two photon radiation in $W$ and $Z$ boson
production~\cite{BS} indicates that,
in order to measure the $W$ mass with a precision of less than 20~MeV
in a hadron collider environment as foreseen in Run~IIB and at the
LHC, it will be necessary to take into account multi-photon radiation
effects.
Moreover, an improved understanding of the uncertainty
due to PDFs is needed. At the Tevatron the PDFs can be constrained
by a measurement of the $W$ charge asymmetry. The estimated
uncertainty on the
$W$ mass due to PDF uncertainties in Run~I was 15 MeV, which is
expected to improve in Run~II.

\noindent
Given the estimated time scale of about 6--8 years until Run~IIB
will be completed, it seems reasonable to hope for a
considerable improvement of the intrinsic theoretical
uncertainties of the EWPO, which would arise from full results
at the two-loop level and improved predictions for the dominant
higher-order corrections. The parametric uncertainty induced by
the experimental error of $m_t$ will be further reduced at
Run~IIB, but will still play an important role in the indirect
determination of the Higgs boson mass.

\item \ul{Tevatron Run IIB$^*$ (30 ${\rm fb}^{-1}$/experiment):}\\
If the current Fermilab booster is replaced by a high intensity proton driver,
it is conceivable that an integrated luminosity of 30~fb$^{-1}$ can be
achieved with the Tevatron by 2008--9~\cite{pdriver}. For integrated
luminosities
larger than 15~fb$^{-1}$,
the uncertainty on the $W$ mass extracted using the traditional
transverse mass method is dominated by systematic uncertainties
associated with the production and decay model~\cite{talkmschmitt}. This
uncertainty
can be reduced significantly by using the $W/Z$ transverse mass
ratio~\cite{run2prec} to measure $\MW$. Extrapolating from the present
uncertainties of $\de\MW({\rm stat})=211$~MeV and $\de\MW({\rm
sys})=50$~MeV obtained using the $W/Z$ transverse mass ratio method (see
Ref.~\cite{run2prec} and references therein), one
finds that an overall uncertainty of $\de\MW=10-15$~MeV might be
achievable for an integrated luminosity of 30~fb$^{-1}$.

\item \ul{LHC:}\\
$\MW$ and $\mt$ will be measured using techniques similar to those
employed at the Tevatron. In order to improve the experimental
uncertainty of $\sweff$ at the LHC, it will be necessary to detect one
of the leptons originating from $Z\to l^+l^-$ over the entire pseudorapidity
range of $|\eta|<5$~\cite{haywood}. This requires an electron jet
rejection factor of $<0.01$ in the forward region ($2.5<|\eta|<5$) of the
electromagnetic calorimeter. The relevance of a more precise
determination of PDFs in this respect remains to be investigated.

The improvement in the measurement of $\MW$ at the LHC is due to the
large statistics which is expected to result in very small statistical
errors and good control of many systematic uncertainties. However, as
in Run~IIB, theoretical improvements are needed, e.g.\ for radiative
$W$ decays, the modeling of the $p_T^W$ distribution, and for
constraining PDFs.  In Ref.~\cite{haywood} it has been argued that it
should be possible to obtain an uncertainty on the $W$ mass due to
PDFs smaller than 10 MeV.
\item \ul{LC (without GigaZ option):}\\
As for the $\MW$ measurement
at LEP2, the determination of the $W$ mass at the LC at center of mass (CM)
energies above the $W^+W^-$ production threshold will be based on
direct reconstruction of
$W$-pair events in 4-fermion production processes. The small
experimental uncertainty at LEP2 and the LC requires the inclusion of
electroweak radiative corrections to the predictions for the
underlying production processes, $e^+e^- \to WW \to 4f$.  The full
treatment of the processes $e^+e^- \to 4f$ at the one-loop level is of
enormous complexity.  Nevertheless, there is ongoing work in this
direction \cite{Vicini:1998iy,vicini}.  While the real Bremsstrahlung
contribution is known exactly, there are severe theoretical problems
with the virtual $O(\alpha)$ corrections.  A detailed description of
the status of predictions for $e^+e^- \to 4f$ processes can be found
in \cite{mclep2}.  A suitable approach to include $O(\alpha)$
corrections to gauge-boson pair production is a double-pole
approximation (DPA): electroweak ${\cal O}(\alpha)$ corrections are
only considered for the terms that are enhanced by two resonant gauge
bosons. All present calculations of $O(\alpha)$ corrections to
$e^+e^-\to WW \to 4f$ rely on a DPA
\cite{Beenakker_gr,Denner_bj,Jadach_hi,Jadach_tz,Kurihara_ii},
and different versions of a DPA have been implemented in the
state-of-the-art Monte Carlo (MC) generators {\tt RacoonWW}
\cite{Denner_dt,Denner_kn,Denner_bj} and {\tt YFSWW3}
\cite{Jadach_hi,Jadach_tz,Jadach_uu}.  The intrinsic DPA error
is estimated to be of the order of $\alpha \Gamma_W/(\pi M_W)$,
i.e.~${\;\raisebox{-.3em}{$\stackrel{\displaystyle <}{\sim}$}\;}$~0.5\%,
whenever the cross section is dominated by doubly-resonant
contributions. This is the case at LEP2 for energies sufficiently
above threshold. The DPA is not a valid approximation close to the
$W$-pair production threshold. At higher energies, the contributions of
single resonant and non-resonant diagrams become sizeable,
and appropriate cuts may need to be imposed to extract the $WW$
signal.

An estimate of the theoretical uncertainty of the $\MW$ measurement at
LEP2 due to electroweak corrections when using the state-of-the-art
MC programs has been given in~\cite{Jadach_cz} by exploiting
numerical results obtained at 200~GeV with {\tt KORALW} and {\tt
YFSWW3}. Using idealized event selections and a simple fitting
procedure, the theoretical uncertainty on $\MW$ is estimated to be
about 5 MeV.  In view of an envisioned 10~MeV measurement at the LC in
the continuum this analysis should be repeated using realistic LEP2
event selection criteria, and for the LC CM energy of 500~GeV.

\noindent
At the LC, the top quark mass can either be extracted from a
$t\bar t$ threshold scan that would determine a suitably defined
threshold mass~\cite{tttheo,ttexp}, or in the continuum by direct
kinematical reconstruction of
$e^+e^-\to\bar tt\to W^+W^-b\bar b\to l^+\nu l^-\bar\nu b\bar b$
events~\cite{yeh} which determines the pole mass. The remaining
theoretical uncertainties are sufficiently small to allow a
measurement of the threshold mass with a precision of ${\cal
O}(50$~MeV)~\cite{tttheo,ttexp,mtnew}.
The measurement of the pole mass at higher energies
with an accuracy of 200 MeV or better may be possible~\cite{yeh},
but is limited in precision
by QCD renormalon effects which are of ${\cal O}(\Lambda_{QCD})$.

The precise measurement of $m_t$ at the LC will eliminate the
main source of parametric uncertainties of the EWPO. The
uncertainties induced in $\MW$ and $\sweff$ by the experimental
error of $m_t$ will be reduced by the LC measurement to the
level of 1~MeV and $0.5 \times 10^{-5}$, respectively, i.e.\
far below the uncertainty corresponding to the present error of
$\de\De\al_{\rm had}$.

\item \ul{GigaZ:}\\
A determination of $\MW$ with the GigaZ option is
based on a dedicated threshold scan. Presently, the predictions for
the $W^+W^-$ cross section in the threshold region are based on an
improved-Born approximation \cite{Beenakker_kt,Denner_zp}
which neglects non-universal electroweak corrections.  Thus, the total
$W$-pair cross section in the threshold region is only known with an
accuracy of about $1.4\%$~\cite{Beenakker_kt}. This translates
into a theoretical uncertainty on the $W$ mass of about
24~MeV~\cite{run2prec,lep2rep,stirling_1}.  Since the extracted value
for $\MW$ may be more sensitive to the shape of the cross section than
its normalization, it has been suggested that this estimate is too
pessimistic, neglecting possible cancellations in cross section
differences.  However, as discussed in more detail in Ref.~\cite{theoprec},
it is expected that the non-universal corrections noticeably affect
the shape in the threshold region.  Thus, in order to achieve the
target precision of $\de\MW=7$~MeV, a full \order{\al} calculation of
the process $e^+e^- \to WW \to 4f (+\gamma)$ in the threshold region
is needed. This is a very difficult task, in particular since
currently no practicable solution of the gauge invariance problem
associated with finite $W$-width effects in loop calculations exists.
Aiming at an accuracy of $\de\MW \approx 7$~MeV will clearly require
a considerable effort from the theory side.
Besides an estimated future primordial theoretical uncertainty
of $\sim 2-3$~MeV, the experimental error for $\MW$
also includes the uncertainty arising from the beam energy.
It has to be controlled at the level of $\sim 2.5$~MeV, which,
although it is of higher precision as currently foreseen for TESLA
or NLC, might be achievable with some additional effort~\cite{wilson}.

\noindent
At GigaZ one hopes to improve the current precision of $\sweff$ by
more than an order of magnitude.
This is envisaged by a precise measurement of
$A_{\rm LR}$~\cite{moenig,sn_pol} using the Blondel scheme~\cite{blondel}.
$A_{\rm LR}$ is then given as a function of polarized cross sections,
where both beams have different combinations of polarizations.
Due to the anticipated drastic improvement in the accuracy, a
reanalysis of the effect of primordial uncertainties in the
determination of $\sweff$ might become necessary.
This determination of $A_{\rm LR}$ requires that both beams can be
polarized independently and that
the polarizations of the colliding $e^+$~and $e^-$~bunches with
opposite helicity states are equal (or that their difference is
precisely determined; see Ref.~\cite{moenig} for details). A precision of
$\de A_{\rm LR} \approx 8 \times 10^{-5}$ seems to be
feasible~\cite{moenig,teslatdr}, resulting in
$\de\sweff \approx 10^{-5}$.

\noindent
The $t \bar t$ threshold analysis at the LC will result in correlated
measurements of $\als$ and $\mt$. Since an independent
and more precise determination of $\als(\MZ)$ would be possible with GigaZ 
(to $\pm 0.0010$, from other GigaZ
observables: the $Z$~width with 1~MeV uncertainty, and $R_l$ with 0.05\%
uncertainty~\cite{jenssvengigaz,alphas}), an
improved value for $\mt$ can be expected as well.

In view of the increased precision of $\sweff$ at GigaZ and the
largely reduced error of $\mt$ at the LC, it will be
very important to reduce the
uncertainty of $\de\De\alpha_{\rm had}$ at least to the level of
$\de\Delta\alpha (M_Z) = 7\times 10^{-5}$~\cite{jegerlehner},
corresponding to parametric uncertainties of $\MW$ and $\sweff$
of $1.5$~MeV and $2.5 \times 10^{-5}$, respectively.
This will require improved measurements of
$R \equiv \sigma(e^+e^- \to {\rm hadrons})/
          \sigma(e^+e^- \to \mu^+\mu^-)$
below about $\sqrt{s} \le 5$~GeV.
In case the uncertainty of $\Delta\alpha (M_Z)$
could even be improved by another factor of two (and taking also
into account the expected improvement in the $\als(M_Z)$
determination at GigaZ), the limiting factor in the parametric
uncertainty of $\sweff$ would arise from the experimental error
of $M_Z$ ($\de M_Z = 2.1$~MeV induces an uncertainty of
$1.4 \times 10^{-5}$ in $\sweff$), which is not expected to
improve in the foreseeable future.

With the prospective future improvements of higher order
corrections to the EWPO discussed above (i.e.\ complete
electroweak two-loop results and a reduction of the
uncertainties in the lower part of Table~\ref{higherorderunc}
by a factor of two), the intrinsic theoretical uncertainties
of the EWPO will be comparable to or smaller than the parametric
uncertainties and the experimental errors at GigaZ (see also
Ref.~\cite{gigaztheo}.)

In summary, the projected experimental accuracies at
GigaZ require on the theory side a 
considerable effort to reduce primordial
theoretical uncertainties. In addition, improvements of the
intrinsic and parametric uncertainties of the EWPO are needed. These tasks
appear challenging, but, in view of the time scale of at least a
decade, not unrealistic.

\end{itemize}

Based on the uncertainties expected at
the next generation of colliders and our estimates of
present and future theoretical uncertainties, we list in
Table~\ref{indirectmh} the (cumulative) precision of $\MH$ at different
colliders which one hopes to achieve from EWPOs.
Results are given for $\delta \MH/\MH$ obtained from $\MW$
alone, from $\sweff$ alone, and from all precision data, taking into
account the intrinsic and the parametric theoretical uncertainties and
their correlated effects.

\begin{table}[thb]
\caption{The expected {\sl cumulative\/} precision, $\de\MH/\MH$, from future 
collider data, given the error projections in
Tables~\ref{expfuture} and~\ref{higherorderunc}. Intrinsic theoretical 
and parametric uncertainties and their correlated effects
on $\MW$ and $\sweff$ are taken into account.
In the first row, our estimate for the current intrinsic
uncertainties in $\MW$ and $\sweff$ from unknown higher order
corrections as given in Table~\ref{higherorderunc} is used.
In the other rows we assume that complete two-loop results for
the most relevant EWPO are available, and that the
uncertainties in the lower part of Table~\ref{higherorderunc}
have been reduced by a factor of two. This corresponds to
future intrinsic theoretical uncertainties in $\MW$ and $\sweff$
of 3~MeV and $1.7\times 10^{-5}$, respectively$^a$.
As in Eq.~(\ref{eq:presentfit}) we have added the theoretical and
experimental errors in quadrature. We also assume
$\de\Delta\alpha (M_Z) = 7\times 10^{-5}$~\cite{jegerlehner}.
(Using the very optimistic value of $5\times 10^{-5}$ would improve
the $\de\MH$ uncertainty at GigaZ to 7\%.)
The last row also assumes a determination of $\alpha_s (M_Z)$
with an uncertainty of $\pm 0.0010$ from other GigaZ observables. }
\label{indirectmh}
\renewcommand{\arraystretch}{1.5}
\begin{tabular}{|l||r|r||r|}

\hline
$\delta\MH/\MH$ {\it from:}            & $\MW$ & $\sweff$ & all  \\
\hline \hline
now                                    &106 \% & 60 \%    & ~58 \% \\
\hline \hline
Tevatron Run IIA                       & 72 \% & 39 \%    & ~35 \% \\
\hline
Tevatron Run IIB                       & 37 \% & 33 \%    & ~25 \% \\
\hline
Tevatron Run IIB$^*$                   & 30 \% & 29 \%    & ~23 \% \\
\hline
LHC                                    & 22 \% & 25 \%    & ~18 \% \\
\hline \hline
LC                                     & 15 \% & 24 \%    & ~14 \% \\
\hline
GigaZ                                  & 12 \% &  8 \%    &  ~8 \% \\
\hline
\end{tabular}
\renewcommand{\arraystretch}{1}
\end{table}

\renewcommand{\thefootnote}{\alph{footnote}}
\footnotetext[1]{Technically within {\tt GAPP} this is realized by
assuming an uncertainty in the $T$ parameter, $T = 0 \pm 0.007$.}
If the SM is the correct low energy theory, the Higgs boson will be
discovered in Run~II at the Tevatron or at the LHC. In this case, the
indirect determination of $\MH$ from precision electroweak measurements
will constitute an important internal
consistency check of the SM. Possible new scales beyond the SM could
manifest themselves in a disagreement of the directly and
indirectly
determined $\MH$ value~\cite{jenssvengigaz,sitgesproc}.


\begin{acknowledgments}

U.B. is supported by NSF grant PHY-9970703.
The work of D.W. is supported by the U.S. Department
of Energy under grant DE-FG02-91ER40685. 
D.R.W. is supported by NSF grant PHY-9972170.

\end{acknowledgments}

\bibliography{BlueBand}

\end{document}